# FRACTIONAL WKB APPROXIMATION

Eqab M. Rabei∗ and Ibrahim M.A.Altarazi
Department of physics, Mu'tah University, Al-karak, Jordan
Sami I. Muslih
Department of physics, Al-Azhar University, Gaza, Palestine
Dumitru Baleanu
Department of Mathematics and Computer Science, Faculty of Arts and Science, Çankaya University
and
Institute of Space Sciences, P.O.BOX, MG-23, R 76900, Magurele-Bucharest, Romania

**Abstract**

Wentzel, Kramers, Brillouin (WKB) approximation for fractional systems is investigated in this paper using the fractional calculus. In the fractional case the wave function is constructed such that the phase factor is the same as the Hamilton's principle function "S". To demonstrate our proposed approach two examples are investigated in details.

∗eqabrabei@yahoo.com

## I. Introduction

Fractional calculus is a branch of mathematics that deal with a generalization of well-known operations of differentiations and integrations to arbitrary non-integer order, which can be real non-integer or even imaginary number.

Nowadays physicists have used this powerful tool to deal with some problems which were not solvable in the classical sense. Therefore, the fractional calculus became one of the most powerful and widely useful tools in describing and explaining some physical complex systems.

Recently, the Euler-Lagrange equations has been presented for unconstrained and constrained fractional variational problems [1 and other references]. This technique enable us to solve some problems including describing the behavior of non-conservative systems developed by Riewe [2], where he used the fractional derivative to construct the Lagrangian and Hamiltonian for non-conservative systems.

From these reasons in [3] was developed a general formula for the potential of any arbitrary force conservative or not conservative, which leads directly to the consideration of dissipative effect in Lagrangian and Hamiltonian formulation. Also, the canonical quantization of non-conservative systems has been carried out in [4].

Starting from a Lagrangian containing a fractional derivative, the fractional Hamiltonian is achieved in [5]. In addition, the passage from Hamiltonian containing fractional derivatives to the fractional Hamilton-Jaccobi is achieved by Rabei et.al [6]. The equations of motion are obtained in a similar manner to the usual mechanics.

All these outstanding results using the fractional derivative make us concentrate on another branch of quntam physics. WKB approximation [7, 8, 9, 10,14]. In this paper we are mainly interested to construct the solution of Schroödinger equation in an exponential form (Griffith 1995) starting from fractional Hamilton-Jaccobi equation and how it leads naturally to this semi-classical approximation namely fractional WKB.

The purpose of this paper is to find the solution of Schrödinger equation for some systems that have a fractional behavior in their Lagrangians and obey the WKB approximation assumptions.

The plan of this paper is as follows: In section **II** the derivation of generalized Hamilton-Jaccobi partial differential equation which given in [6] is briefly reviewed. In section **III** the fractional WKB approximation is derived. In Section **IV** some examples with the fractional WKB technique is reported. Section **V** is dedicated to conclusions.

**II. Basic Tools**

The left and right Reimann-Loville fractional derivative are defined as follows [3]

The left Riemann-Liouville fractional derivative is given by

$$_aD_x^\alpha f(x) = \frac{1}{\Gamma(n-\alpha)}\left(\frac{d}{dx}\right)^n \int_a^x (x-\tau)^{n-\alpha-1} f(\tau)\, d\tau \qquad (1)$$

The right Riemann-Liouville fractional derivative has the form

$$_xD_b^\beta f(x) = \frac{1}{\Gamma(n-\alpha)}\left(-\frac{d}{dx}\right)^n \int_x^b (\tau-x)^{n-\beta-1} f(\tau)\, d\tau \qquad (2)$$

Here α, β are the order of derivation such that n-1≤α <n, n-1≤β<n, and they are not zero.

If α is an integer, these derivatives are defined in usual sense as

$$_aD_x^\alpha f(x) = \left(\frac{d}{dx}\right)^\alpha f(x) \qquad (3\text{-a})$$

$$_xD_b^\beta f(x) = \left(-\frac{d}{dx}\right)^\beta f(x) \qquad (3\text{-b})$$

Hamilton formalism with fractional derivative was proposed in [5] namely

$$H(q, p_\alpha, p_\beta, t) = p_\alpha\, _aD_t^\alpha q + p_\beta\, _tD_b^\beta q - L(q,\, _aD_t^\alpha q,\, _tD_b^\beta q, t)\;, \qquad (4)$$

where L represents the fractional Lagrangian obtained by replacing the classical derivatives with the corresponding fractional ones [5].

Hamilton's equations of motion are obtained as follows [5]

$$\frac{\partial H}{\partial t} = -\frac{\partial L}{\partial t}; \quad \frac{\partial H}{\partial p_\alpha} = {}_aD_t^\alpha q; \quad \frac{\partial H}{\partial p_\beta} = {}_tD_b^\beta q; \quad \frac{\partial H}{\partial q} = {}_aD_t^\beta p_\beta + {}_tD_b^\alpha p_\alpha \qquad (5)$$

In [6] based on the sequential derivatives the fractional Hamilton-Jacobi partial differential equation is obtained. The Hamilton-Jacobi function in configuration space is written in a similar manner to the usual mechanics by using the Reimann-Loville fractional derivative. In [6] the following generating function $F = F_2({}_aD_t^{\alpha-1}q, {}_tD_b^{\beta-1}q, P_\alpha, P_\beta, t) = S$ is used, where α and β are bigger or equal to 1. Thus, the new Hamiltonian is expressed as

$$K(Q, P_\alpha, P_\beta, t) = P_\alpha \; {}_aD_t^\alpha Q + P_\beta \; {}_tD_b^\beta Q - L'(Q, {}_aD_t^\alpha Q, {}_tD_b^\beta Q, t) \qquad (7)$$

It is concluded that, the following relation relates the two Hamiltonians

$$p_\alpha \; {}_aD_t^\alpha q + p_\beta \; {}_tD_b^\beta q - H = P_\alpha \; {}_aD_t^\alpha Q + P_\beta \; {}_tD_b^\beta Q - K + \frac{dF}{dt} \qquad (8)$$

According to reference [6] the function F is proposed as

$$F = S({}_aD_t^{\alpha-1}q, {}_tD_b^{\beta-1}q, P_\alpha, P_\beta, t) - P_\alpha \; {}_aD_t^{\alpha-1}Q - P_\beta \; {}_tD_B^{\beta-1}Q, \qquad (9)$$

The function S is called Hamilton's principle function.

Therefore, requiring that the transformed Hamiltonian $K$ shall be zero the Hamilton-Jacobi equation is satisfied. In other words Q, $P_\alpha$, $P_\beta$ are constants.

$$H + \frac{\partial S}{\partial t} = 0 \qquad (10)$$

Since Q, $P_\alpha$, $P_\beta$ are constants, The Hamilton's principle function is written as

$$S = S({}_aD_t^{\alpha-1}q, {}_tD_b^{\beta-1}q, E_1, E_2, t) \qquad (11)$$

where

$$P_\alpha = E_1 \qquad P_\beta = E_2$$

If the Hamiltonian is explicitly independent of time, then S can be written as follows

$$S = W_1({}_aD_t^{\alpha-1}q, E_1) + W_2({}_tD_b^{\beta-1}q, E_2) + f(E_1, E_2, t) \tag{12}$$

W represents the Hamilton's characteristic function; therefore, the following equations of motion are obtained in [6] as:

$$P_\alpha = \frac{\partial W_1}{\partial {}_aD_t^{\alpha-1}q} \qquad P_\beta = \frac{\partial W_1}{\partial {}_tD_b^{\beta-1}q} \tag{13}$$

$$_aD_t^{\alpha-1}Q = \frac{\partial W_1}{\partial E_1} = \lambda_1 \qquad {}_tD_b^{\beta-1}Q = \frac{\partial W_2}{\partial E_2} = \lambda_2 \tag{14}$$

Here $\lambda_1, \lambda_2$ are constants.

### III. Fractional WKB approximation

The outstanding result regarding the meaning of the state function $\psi$ and its relationship to Hamilton's principle function S enables us to write the exponential solution of Schrödinger equation [13].

$$\psi(q,t) = \exp\left(\frac{iS(q,t)}{\hbar}\right) \tag{15}$$

The phase of state function obeys the same mathematical equation, as does Hamilton's principle function S. The physical significance of S in classical mechanics is that it represents the generator of trajectories [12] for fractional systems; the fractional Hamilton's principle function is become the phase of the state function $\psi$. One can write the solution of Schrödinger equation under the postulated constrains by the WKB approximation and using the fractional Hamilton's principle function eq (12). Thus we propose the fractional state function as:

$$\psi({}_aD_t^{\alpha-1}q, {}_tD_b^{\beta-1}q, t) = \exp\left(\frac{i}{\hbar}S({}_aD_t^{\alpha-1}q, {}_tD_b^{\beta-1}q, t)\right) \tag{16}$$

From the quantization using WKB approximation [7,8,9,10,14] a general solution of Schrödinger equation is obtained using the expansion for S and then using the transformation to the N-dimensional system as:

$$\psi = \left[\prod_{i=1}^{N}\psi_{io}(q_i)\right]\exp\left(\frac{iS(q_i,t)}{\hbar}\right) \tag{17}$$

where

$$\psi_{io}(q_i) = \frac{1}{\sqrt{p(q_i)}} \tag{18}$$

In our case, S behaves like a 2-dimensional problem with two distinct momenta. Thus,

$$q_1 \equiv {}_aD_t^{\alpha-1}q \qquad P_1 \equiv \hat{P}_\alpha \tag{19}$$

$$q_2 \equiv {}_tD_b^{\beta-1}q \qquad P_2 \equiv \hat{P}_\beta \tag{20}$$

And the momenta are defined as operators. Therefore, we can propose the wave function $\psi$ of the fractional system in the following form

$$\psi({}_aD_t^{\alpha-1}q, {}_tD_b^{\beta-1}q, t) = \frac{1}{\sqrt{P_\alpha P_\beta}}\exp\left(\frac{i}{\hbar}S({}_aD_t^{\alpha-1}q, {}_tD_b^{\beta-1}q, E_1, E_2, t)\right) \tag{21}$$

and the momenta operators in the form

$$\hat{P}_\alpha = \frac{\hbar}{i}\left(\frac{\partial}{\partial\, {}_aD_t^{\alpha-1}q}\right), \quad \hat{P}_\beta = \frac{\hbar}{i}\left(\frac{\partial}{\partial\, {}_tD_b^{\beta-1}q}\right) \tag{22}$$

We conclude that (21) is the solution of Schrödinger equation for any given fractional systems. If α and β both are equal to unity, then we will return to the usual classical solution of Schrödinger equation, also we can notice how the probability is inversely proportional to the momentum

$$|\psi|^2 \cong \frac{1}{p(q)}.$$

## IV. Examples

### IV. a) Example 1:

As a first model let us consider the following fractional Lagrangian,

$$L = \frac{1}{2}\left({}_0D_t^\alpha q\right)^2 + \frac{1}{2}\left({}_tD_1^\beta q\right)^2 \qquad (23)$$

The fractional Hamilton-Jacobi equation for this fractional Lagrangian can be calculated as:

$$\frac{1}{2}(P_\alpha)^2 + \frac{1}{2}(P_\beta)^2 + \frac{\partial S}{\partial t} = 0. \qquad (24)$$

where

$$P_\alpha = \frac{\partial L}{\partial {}_0D_t^\alpha q} \quad ; \quad P_\beta = \frac{\partial L}{\partial {}_tD_1^\beta q}$$

Making use of equation (13), the fractional Hamilton-Jacobi equation (24) becomes:

$$\frac{1}{2}\left(\frac{\partial W_1}{\partial {}_0D_t^{\alpha-1} q}\right)^2 + \frac{1}{2}\left(\frac{\partial W_2}{\partial {}_tD_1^{\beta-1} q}\right)^2 + \frac{\partial S}{\partial t} = 0 \qquad (25)$$

Taking into account

$$H = -\frac{\partial S}{\partial t} \qquad (26)$$

If we apply (26) on a wave function it gives:

$$\frac{\partial S}{\partial t} = -E \equiv -(E_1 + E_2) \qquad (27)$$

By using the fact that E is the total energy of the system and taking into account (27) we obtain

$$\left[\frac{1}{2}\left(\frac{\partial W_1}{\partial\,_0D_t^{\alpha-1}q}\right)^2 - E_1\right] + \left[\frac{1}{2}\left(\frac{\partial W_2}{\partial\,_tD_1^{\beta-1}q}\right)^2 - E_2\right] = 0 \qquad (28)$$

Thus, both sides of (28) should be zero, and we obtain

$$W_1 = \sqrt{2E_1}\,_0D_t^{\alpha-1}q, \qquad\qquad W_2 = \sqrt{2E_2}\,_tD_1^{\beta-1}q \qquad (29)$$

By using (12) and (21) we obtain

$$\psi(_0D_t^{\alpha-1}q,\,_tD_1^{\beta-1}q,t) = \frac{1}{\sqrt{P_\alpha P_\beta}}\exp\left(\frac{i}{\hbar}\left(\sqrt{2E_1}\,_0D_t^{\alpha-1}q + \sqrt{2E_2}\,_tD_1^{\beta-1}q - Et\right)\right)$$

(30)

Which represents the wave function of the following Hamiltonian:

$$H = \frac{1}{2}(P_\alpha)^2 + \frac{1}{2}(P_\beta)^2 \qquad (31)$$

Let us deal now with the momenta as operators of the form (22), and applying these operators on the wave function, one obtain the following momenta eigenvalues

$$\hat{P}_\alpha\psi = \sqrt{2E_1}\,\psi \qquad\qquad \hat{P}_\beta\psi = \sqrt{2E_2}\,\psi \qquad (32)$$

Then,

$$\left|\hat{P}_\alpha\right| = \sqrt{2E_1} \qquad\qquad \left|\hat{P}_\beta\right| = \sqrt{2E_2} \qquad (33)$$

It's the same as the classical solution. Also, when applying the energy operator it gives the energy eigenvalues:

$$H\psi = \frac{1}{2}(\hat{P}_\alpha)^2\psi + \frac{1}{2}(\hat{P}_\beta)^2\psi \quad = (E_1 + E_2)\psi \qquad (34)$$

as in the classical case.

**IVb.) Example 2:**

As a second example let us consider the following fractional Lagrangian

$$L = \frac{1}{2}\left(_0D_t^\alpha q\right)^2 + \frac{1}{2}\left(_tD_1^\beta q\right)^2 + {_0D_t^\alpha q} + {_tD_1^\beta q} + \frac{1}{2}q^2 \qquad (35)$$

The corresponding fractional Hamilton is calculated as follows

$$H = \frac{1}{2}\left(\hat{P}_\alpha - 1\right)^2 + \frac{1}{2}\left(\hat{P}_\beta - 1\right)^2 - \frac{1}{2}q^2 \qquad (36)$$

Thus, the fractional Hamilton-Jacobi equation becomes

$$\frac{1}{2}\left(\hat{P}_\alpha - 1\right)^2 + \frac{1}{2}\left(\hat{P}_\beta - 1\right)^2 - \frac{1}{2}q^2 + \frac{\partial S}{\partial t} = 0 \qquad (37)$$

The fractional Hamilton's principle function is calculated as,

$$S = \left(\sqrt{q^2 + 2E_1} + 1\right){_0D_t^{\alpha-1}q} + \left(\sqrt{2E_2} + 1\right){_tD_1^{\beta-1}q} - (E_1 + E_2)t \qquad (38)$$

As a result the wave function can be written in the form

$$\psi(_0D_t^{\alpha-1}q, {_tD_1^{\beta-1}q}, t) = \frac{1}{\sqrt{P_\alpha P_\beta}} \exp\left(\frac{i}{\hbar}\left(\left(\sqrt{q^2 + 2E_1} + 1\right){_0D_t^{\alpha-1}q} + \left(\sqrt{2E_2} + 1\right){_tD_1^{\beta-1}q} - Et\right)\right) \quad (39)$$

To identify the influence of the operators let us test the effect of the momenta

$$\hat{P}_\alpha = \frac{\hbar}{i}\left(\frac{\partial}{\partial\, _aD_t^{\alpha-1}q}\right) \qquad \hat{P}_\beta = \frac{\hbar}{i}\left(\frac{\partial}{\partial\, _tD_b^{\beta-1}q}\right) \qquad (40)$$

Using the characteristic equations, it can be shown that

$$\left|\hat{P}_\alpha\right| = \sqrt{q^2 + 2E_1} + 1, \qquad \left|\hat{P}_\beta\right| = \sqrt{2E_2} + 1 \qquad (41)$$

The result shown in (41) is the same classical solution. When applying the energy operator it will give the energy eigenvalues

$$H\psi = \frac{1}{2}\left(\hat{P}_\alpha - 1\right)^2\psi + \frac{1}{2}\left(\hat{P}_\beta - 1\right)^2\psi - \frac{1}{2}q^2\psi \qquad (42)$$

$$= \frac{1}{2}\left(\hat{P}^2{}_\alpha + \hat{P}^2{}_\beta\right)\psi - \left(\hat{P}_\alpha + \hat{P}_\beta\right)\psi + \psi - \frac{1}{2}q^2\psi$$

$$= \{\frac{1}{2}\left(q^2 + 2E + 2\sqrt{q^2 + 2E_1} + 2\sqrt{2E_2} + 2\right)$$

$$- \left(\sqrt{q^2 + 2E_1} + \sqrt{2E_2} + 2\right) + 1 - \frac{1}{2}q^2\}\psi$$

Then we get

$$H\psi = E\psi \tag{43}$$

which is exactly the total energy as the case for the classical systems.

## V. Conclusions

We use the generating function "S" of the Hamilton-Jaccobi equation in its fractional form to be the phase factor of the wave function describing some potentials valid for the assumptions suggested by the WKB approximation

The proof of our results arises from the new proposed concepts of the momentum and energy operators, that they give the same eigenvalues producing the ordinary results achieved by the classical approach.

Giving the same eigenvalues that means this form of fractional operator also eigen, valid, and useful in effecting on a state functions.